\documentclass[%
 reprint,
superscriptaddress,
showpacs,
 amsmath,amssymb,
 aps,
]{revtex4-1}

\usepackage{graphicx}% Include figure files
\usepackage{dcolumn}% Align table columns on decimal point
\usepackage{bm}% bold math
\begin{document}

\preprint{APS/123-QED}

\title{Angular distributions of the vector $A_y$ and tensor $A_{yy}$, $A_{xx}$, $A_{xz}$ analyzing powers in the $\vec{d}d\to{\rm^3H}p$ reaction at 200 MeV}% Force line breaks with \\

\author{A.K.~Kurilkin}%
 \email{akurilkin@jinr.ru}
 \affiliation{LHEP-JINR, 141-980 Dubna, Moscow region, Russia}%Lines break automatically or can be forced with \\
\author{T.~Saito}%
\affiliation{Faculty of Engineering, University of Miyazaki, Miyazaki, 889-2192, Japan}
\author{V.P.~Ladygin}
 \affiliation{LHEP-JINR, 141-980 Dubna, Moscow region, Russia}
\author{T.~Uesaka}
\affiliation{Center for Nuclear Study, University of Tokyo, Bunkyo, Tokyo 113-0033, Japan}
%\affiliation{RIKEN, Wako, Saitama 351-0198, Japan}
\author{M.~Hatano}
\affiliation{Department of Physics, University of Tokyo, Bunkyo, Tokyo 113-0033, Japan}
\author{A.Yu.~Isupov}
\affiliation{LHEP-JINR, 141-980 Dubna, Moscow region, Russia}
\author{M.~Janek}
\affiliation{Physics Department, University of $\check{Z}$ilina, 01026 $\check{Z}$ilina, Slovakia}
\author{H.~Kato}
\affiliation{Department of Physics, University of Tokyo, Bunkyo, Tokyo 113-0033, Japan}
\author{N.B.~Ladygina}
\affiliation{LHEP-JINR, 141-980 Dubna, Moscow region, Russia}
\author{Y.~Maeda}
%\affiliation{Faculty of Engineering, University of Miyazaki, Miyazaki, 889-2192, Japan}
\affiliation{Kyushi University, 6-10-1 Hakozaki, Higashi-ku, Fukuoka-shi 812, Japan}
\author{A.I.~Malakhov}
\affiliation{LHEP-JINR, 141-980 Dubna, Moscow region, Russia}
\author{J.~Nishikawa}
\affiliation{Department of Physics, Saitama University, UraWa 338-8570, Japan}
\author{T.~Ohnishi}
\affiliation{RIKEN, Wako, Saitama 351-0198, Japan}
\author{H.~Okamura}
\email{Deceased}
\affiliation{Research Center for Nuclear Physics, Osaka University, Ibaraki 567-0047, Japan}
\author{S.G.~Reznikov}
\affiliation{LHEP-JINR, 141-980 Dubna, Moscow region, Russia}
\author{H.~Sakai}
\affiliation{Department of Physics, University of Tokyo, Bunkyo, Tokyo 113-0033, Japan}
\affiliation{Center for Nuclear Study, University of Tokyo, Bunkyo, Tokyo 113-0033, Japan}
%\affiliation{University of Tokyo, Bunkyo, Tokyo 113-0033, Japan}
\author{N.~Sakamoto}
\affiliation{RIKEN, Wako, Saitama 351-0198, Japan}
\author{S.~Sakoda}
\affiliation{Department of Physics, University of Tokyo, Bunkyo, Tokyo 113-0033, Japan}
\author{Y.~Satou}
\affiliation{Research Center for Nuclear Physics, Osaka University, Ibaraki 567-0047, Japan}
\author{K.~Sekiguchi}
\affiliation{Tohoku University, Sendai, J-9808578, Japan}
\author{K.~Suda}
\affiliation{Center for Nuclear Study, University of Tokyo, Bunkyo, Tokyo 113-0033, Japan}
\author{A.~Tamii}
\affiliation{Research Center for Nuclear Physics, Osaka University, Ibaraki 567-0047, Japan}
\author{N.~Uchigashima}
\affiliation{Department of Physics, University of Tokyo, Bunkyo, Tokyo 113-0033, Japan}
\author{T.A.~Vasiliev}
\affiliation{LHEP-JINR, 141-980 Dubna, Moscow region, Russia}
\author{K.~Yako}
\affiliation{Department of Physics, University of Tokyo, Bunkyo, Tokyo 113-0033, Japan}

\date{\today}% It is always \today, today,

\begin{abstract}

A complete set of analyzing powers for the $\overrightarrow{d}d~\rightarrow~{}^3Hp$ reaction at the kinetic beam energy of 200 MeV  has been measured in the full angular range in the c.m. frame. 
The observed signs of the tensor analyzing powers $A_{yy}$, $A_{xx}$, and $A_{xz}$ at forward and backward directions have clearly demonstrated the sensitivity to the ratio of the $D-$ and $S-$wave components of the 
triton and deuteron, respectively.
The new high-precision data are compared with the prediction of the relativistic multiple-scattering model by using standard wave functions of the three-nucleon bound state and of the deuteron.

%The angular distributions of the vector $A_y$ and tensor $A_{yy}$, $A_{xx}$,  $A_{xz}$ analyzing powers  in the $\overrightarrow{d}d~\rightarrow~{}^3Hp$ reaction obtained at Ed=200 MeV in the angular range 0-180 in the c.m.s. are presented. The observed negative sign of the tensor analyzing powers Ayy, Axx and Axz at small angles clearly demonstrate the sensitivity to the ratio of the D- and S-wave component of the three-nucleon bound state wave function. However, the one-nucleon exchange calculations by using the standard wave functions have failed to reproduce the strong variation of the tensor analyzing powers as a function of the angle in the c.m. 

\begin{description}
\item[PACS numbers]
\verb+ 24.70.+s, \verb+ 21.45.+v, \verb+ 25.10.+s.

\end{description}
\end{abstract}

\maketitle

%\tableofcontents

%\section{\label{sec:level1} Introduction}

The most fundamental questions in nuclear physics are related to the properties of the light nuclei and forces which bind nucleons, protons, and neutrons to form the building blocks of matter.
The essential amount of experimental data sensitive to the structure of light nuclei has been accumulated during the last decades. 
The cross section \cite{5,11,12,13,14,15,16} and spin observables, such as analyzing powers \cite{17,18,19,20}, spin correlation coefficients \cite{21}, and polarization transfer coefficients \cite{22,23} are measured for $Nd$ scattering. 
Large discrepancies between the $Nd$ data and theoretical predictions based on the exact solution of the Faddeev equations with modern $NN$ potentials are obtained. These discrepancies are particularly significant in the angular region of the cross-section minimum and at the energy of incoming nucleons above 60 MeV \cite{24}. 
The inclusion of the $2\pi$-exchange three-nucleon forces (3NF) models such as Tucson-Melbourne \cite{TM} or Urbana \cite{UR9} into theoretical calculations has improved the description of the differential-cross-section data. 
%This result has clearly shown the significance of the $3NFs$ in the nucleus. 
However, theoretical calculations with 3NFs still have difficulties in reproducing the data of some spin observables. At higher energies, not only spin observables but also cross sections at backward scattering indicate the deficiencies of the present 3NF models \cite{25,26,27}. 
A more comprehensive analysis of the internal structure of light nuclei and the $3NFs$ effect also requires the study of four-nucleon systems because they have a number of features not found for three nucleons,   
including the existence of excited states, a complicated reaction mechanism, many channels in [3+1] and [2+2] configurations, and much stronger polarization effects. 

A number of experiments which rely on the detailed knowledge of the spin structure of the bound-state wave functions of $A=3$ nuclei, has been performed at intermediate energies. 
The structure of the ${}^3$He nucleus was investigated using the $^3$He$(p,2p)$ and $^3$He$(p,pd)$ reactions at the TRI-University Meson Facility (TRIUMF) \cite{triumf}. 
It has been found that calculations using realistic $2N$ potentials are unable to reproduce the measured nucleon momentum distribution in the region of internal momentum $q >$ 300 MeV/c. 
The precise data sensitive to the short-range spin structure of ${}^3$He were obtained at 197 MeV at the Indiana University Cyclotron Facility (IUCF) \cite{37} up to $q\approx$ 400 MeV/c. The observed polarization of the neutron and proton at zero nucleon momentum in ${}^3$He($P_n = 0.98$ and $P_p = -0.16$, respectively) is in good agreement with the Faddeev calculations \cite{38}. However, at higher momenta there is a discrepancy, which can be explained due to the uncertainty of the theoretical calculations, as well as to large rescattering effects. 
The differential cross section and spin correlation parameter $C_{yy}$ in the $p+{}^3He$ elastic backward scattering have been measured at $E_p$=200, 300 and 400 MeV to study the reaction mechanisms and the validity of the ${}^3$He wave functions based on realistic $NN$ potentials \cite{41}.

   The $dd\to{}^3Hp({}^3Hen)$ process can be used as an effective tool to investigate the structure ${}^3H$ and ${}^3He$ at short distances. 
The analysis of the polarization effects for these reactions in the one-nucleon exchange (ONE) model \cite{42} has shown that the tensor $A_{yy}$, $A_{xx}$, and $A_{xz}$ analyzing powers at the forward and backward angles are related to the $D/S$ ratio of ${}^3H({}^3He)$ and deuteron wave functions, respectively.
The tensor analyzing power $T_{20}$ has been obtained in the $\overrightarrow{d}d\rightarrow{}^3He(0^\circ)n$ $({}^3H(0^\circ)p)$ reactions at the energy of deuteron 140, 200, and 270 MeV at the Institute of Physical and Chemical Research, Japan (RIKEN). 
The calculations of the ONE model qualitatively describe the energy dependence of $T_{20}$ \cite{43}.
% The positive values and behavior of $T_{20}$ \cite{43} depending on the deuteron energy qualitatively agree with the calculations of the ONE model. 
However, the recent data on the tensor $A_{yy}$, $A_{xx}$ and $A_{xz}$ analyzing powers for the $\overrightarrow{d}d\rightarrow{}^3Hen$ reaction obtained at 270 MeV \cite{44}, have demonstrated a strong disagreement with the predictions of ONE at the angles larger than $15^\circ$ in the c.m. frame.   
This difference may indicate that the knowledge of a short-range structure of ${}^3He$ is not complete. The other reason may be the importance of the addition-to-ONE reaction mechanisms. 
Recently, a formalism based on the Alt-Grassenberger-Sandhas \cite{ags} equations for the four-body case has been developed to describe the data both of the differential cross section and of polarization observables in $\overrightarrow{d}d\rightarrow{}^3Hp({}^3Hen)$ reactions at the deuteron kinetic energies of a few hundred MeV \cite{48,Ladygina_ISHEPP2012}. 
A reasonable agreement between the data and theoretical results has been obtained at the energy of 300 MeV. 
It has been observed that the inclusion of a single-scattering diagram in addition to the ONE mechanism significantly improves the description of the experimental data; however, it is not enough to reproduce the magnitude of the cross section and tensor analyzing power $T_{20}$ \cite{48}.

This paper presents the precise experimental data on the  vector $A_y$ and tensor $A_{yy}$, $A_{xx}$, $A_{xz}$ analyzing powers in the $\overrightarrow{d}d~\rightarrow{}~^3Hp$ reaction at 200 MeV in the full angular range in the c.m. 
The goal of these measurements was to obtain data  sensitive to the $^3H$ spin structure at short distances.  
%The experiment and details of the data processing are described in Sect.2. 
%The obtained results and their comparison with the theoretical calculations  
%are discussed in Sect.3.  The conclusions are given in Sect.4.

%\section{\label{sec:level2} Experiment}

Measurements of the analyzing powers in the $\overrightarrow{d}d~\rightarrow{}~^3Hp$ reaction at 200 MeV have been performed at the RIKEN Accelerator Research Facility (RARF). 
The details of the experiment are discussed elsewhere \cite{43,44}; below, we briefly describe the main items of the experimental procedure.

The high-intensity polarized deuteron beam was produced by the polarized ion source(PIS)\cite{45} and accelerated by the azimutaly varying lield (AVF) and ring cyclotrons up to the energy of 200 MeV. 
The direction of the symmetry axis of the beam polarization was controlled with a Wien filter located at the exit of the PIS. 
The magnitudes of the beam polarization were determined by the Swinger (SwingerPOL) and Droom (DroomPOL) polarimeters, based on the measurement of  asymmetry in the $\overrightarrow{d}p$ elastic scattering with the known large values of the tensor and vector analyzing powers \cite{14,15}.
The Droom polarimeter was used for monitoring the polarization while taking data.  The Swinger polarimeter measured the polarization before and after each run.
The polarization values obtained from both polarimeters agreed with each other within the statistical accuracy; therefore, the beam polarization for each polarization state of the PIS was taken as a weighted average of the values obtained by these polarimeters.  

In the present experiment, the data have been taken with polarized and unpolarized beams for different combinations of the incoming polarization given in terms of the theoretical maximum polarization values ($p_z$,$p_{zz}$) = (0,0), (0,-2), (-2/3,0) and (1/3,1). 
These polarization modes were changed cyclically every 5 seconds by switching the RF transition units of the polarized-ion sources. 
The actual values of the beam polarization were between 32$\%$ and 75$\%$ of the maximum theoretical value. The systematic error due to the uncertainties of the values of the $dp$ elastic scattering analyzing powers does not exceed $\approx 2\%$ both for the vector and tensor polarization of the beam. The systematic and statistical errors have been added in quadrature to calculate the total error of the beam-polarization values.
 
The measurements of the scattered-particle momenta and separation from the primary beam were performed by the magnetic system of the SMART (Swinger and Magnetic Analyzer with a Rotation and a Twister) \cite{46} spectrograph consisting of two dipole and three quadrupole magnets (Q-Q-D-Q-D). 
The detection system of SMART consisted of three plastic scintillation counters and a multiwire drift chamber(MWDC). 
The coincidence of the signal outputs of all the three scintillation counters was employed as the event trigger. 
Pulse heights of the plastic scintillation counters were used to select the particle of interest at the trigger level.
The identification of the scattered particles (${}^3He$, ${}^3H$ and $p$) was based on the energy losses in the plastic scintillators and time-of-flight measurements between the target and detection point.
The distance between the target and the detection point was about 17m, which was enough to separate ${}^3H$, deuterons, and protons with the same momentum.
The MWDC information was taken to reconstruct the particle trajectories in the focal plane. The trajectories of the detected particles at the second focal plane were determined by the least-squares method using the position information obtained from the MWDC. The typical track reconstruction efficiency of the MWDC was better than 99$\%$.
The ion-optical parameters of the SMART spectrograph were also taken into account to calculate the momentum of the particle and emission angle in the target to obtain the track information. 
The resulting energy resolution was $\sim$ 300 keV.

\begin{figure}[t]
\includegraphics[width=88mm,height=140mm]{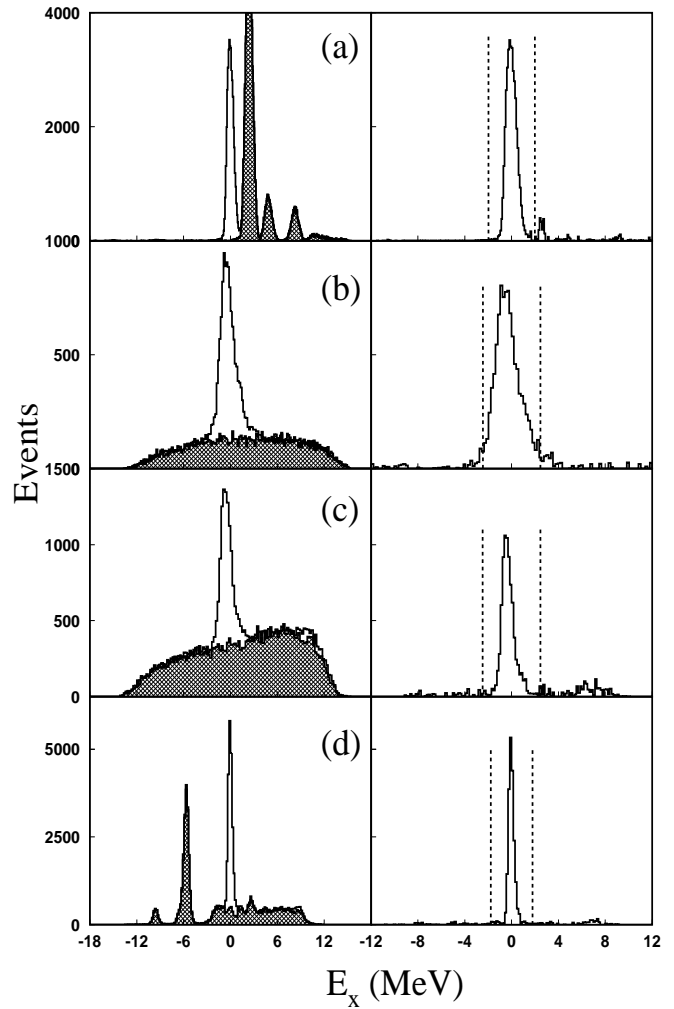}% Here is how to import EPS art
\caption{\label{fig:akurilkin_cd2-c} The $CD_2-C$ subtraction for the $\vec{d}d\rightarrow{}^3Hp$ reaction at $T_d=$ 200 MeV. The open and shadowed histograms in the left panels
correspond to the yields from the $CD_2$ and carbon targets, respectively. The right panels demonstrate the quality of the $CD_2-C$ subtraction.
The panels (a), (b), (c), and (d) correspond to the ${}^3H$ scattering angles in the c.m. frame of $12^{\circ}$, $56^{\circ}$, $144^{\circ}$, and $168^{\circ}$, respectively.
}
\end{figure}

The deuterated polyethylene ($CD_2$) sheet 54 $mg/cm^2$ thick \cite{47} placed in the scattering chamber of the SMART was used as the deuterium target. 
The carbon foil 34 $mg/cm^2$ thick was taken to measure the background spectra. 
The experiment was performed in such a way that only one secondary particle was detected. Tritons and protons were registered in the $0^{\circ}-90^{\circ}$ and $90^{\circ}-180^{\circ}$ angular ranges in the c.m. frame, respectively. 
The contribution of the deuterium target was obtained via the $CD_2-C$ subtraction procedure for each spin state at every angle. 
The subtraction procedure shown in Fig.~\ref{fig:akurilkin_cd2-c} for the $12^{\circ}$, $56^{\circ}$, $144^{\circ}$ and $168^{\circ}$ scattering angle in the c.m. frame.
Figures ~\ref{fig:akurilkin_cd2-c}(a), ~\ref{fig:akurilkin_cd2-c}(b), and ~\ref{fig:akurilkin_cd2-c}(c), ~\ref{fig:akurilkin_cd2-c}(d), correspond to the cases when tritons and protons were registered, respectively.   
The spectra are plotted as a function of the excitation energy $E_x$, which is defined as follows: 
\begin{eqnarray}
E_x = \sqrt{(E_0-E_{})^2-({\bf P}_0-{\bf P}_{})^2} - M, \nonumber
\end{eqnarray}
where ${\bf P}_0$ is the incident momentum, $E_0 = 2M_d + T_d$ is the total initial energy, $E_{}$ and  ${\bf P}_{}$ are the energy and momentum of the registered particle, respectively, and $M$ is the mass of the nonregistered particle. 
%The experiment was performed in the such a way that only one secondary particle was detected. $^3H$ and proton were registered in the $0^{\circ}-90^{\circ}$ and $90^{\circ}-180^{\circ}$ angular ranges in the c.m.s., respectively.  
The left panels represent the relative yields from the $CD_2$ and carbon targets shown by the open and shadowed histograms, respectively. 
The histograms are normalized for the sake of comparison. Peaks at $E_x=0$ MeV correspond to the $\overrightarrow{d}d~\rightarrow~{}^3Hp$ reaction. 
The right panels show the spectra after subtraction of the carbon events normalized to luminosity and corrected for dead time. 
It is clearly demonstrated that the subtraction procedure has been carried out properly. 

The analyzing powers $A_y$, $A_{yy}$, $A_{xx}$ and $A_{xz}$ in the $\overrightarrow{d}d~\rightarrow~{}^3Hp$ reaction were obtained from the number of the events after the $CD_2-C$ subtraction procedure and beam polarization. The number of the events was normalized to the dead-time effect, the detection efficiency, and beam intensity.
Since the polarization modes were cycled every five seconds, the systematic uncertainty due to any time-dependent effects such as deuterium losses from the $CD_2$ target caused by beam irradiation, can be neglected.
%When the angle of the scattered particle in the c.m. was less than 6$^{\circ}$ or larger than 174$^{\circ}$, the azimuthal angle to cover the scattered particles became larger. 
When the angle of the scattered particle in the c.m. frame was less than 6$^{\circ}$ or larger than 174$^{\circ}$, the azimuthal angle covered by the particle detector became between 120$^{\circ}$ and 360$^{\circ}$ depending on the scattering angle. 
In this case, the range of the azimuthal angle was divided into bins of 15$^{\circ}$. The asymmetry from each bin for each polarized spin mode of PIS was acquired individually and the analyzing powers were obtained from the fit of the asymmetry distributions by the function depending on the azimuthal angle. 
   
%\section{Results}

\begin{figure}[ht]
\includegraphics[width=88mm,height=155mm]{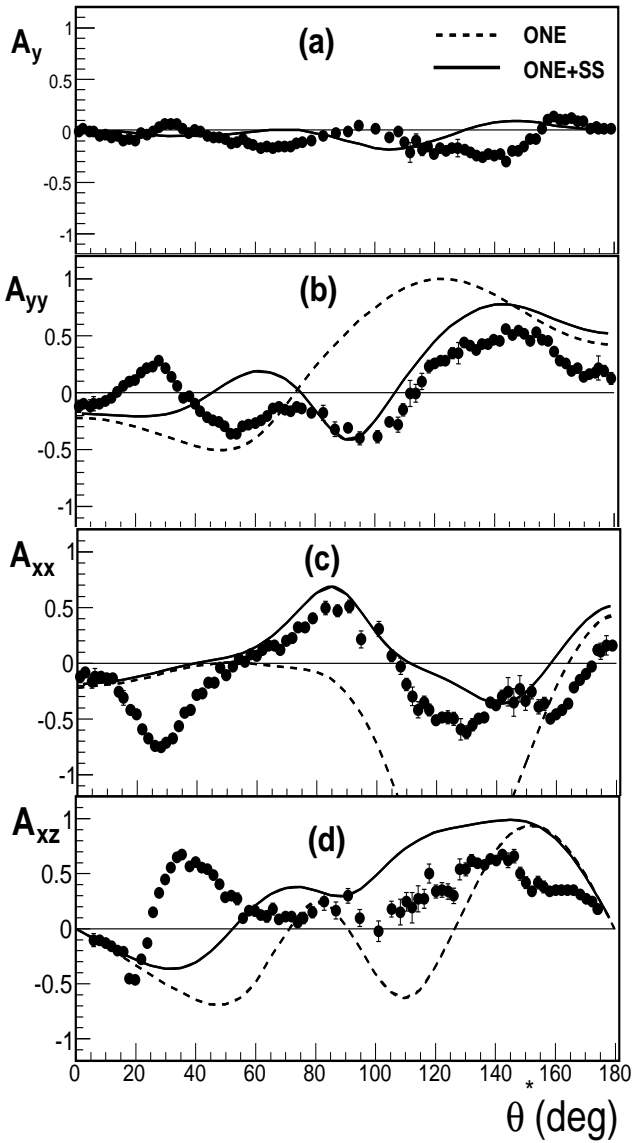}% Here is how to import EPS art
\caption{\label{fig:akurilkin_data200} Results for the vector $A_{y}$ and tensor $A_{yy}$, $A_{xx}$, $A_{xz}$ analyzing powers in the $\vec{d}d\rightarrow{}^3Hp$ reaction at 200 MeV. The dashed and solid lines are theoretical calculations taking into account the ONE and the ONE + SS contributions, respectively. }
\end{figure}

The experimental results on the angular distributions of the vector $A_{y}$ and tensor $A_{yy}$, $A_{xx}$, $A_{xz}$ analyzing powers of the $\overrightarrow{d}d\rightarrow{}^3Hp$ reaction in the c.m. frame at the energy of 200 MeV are presented in Fig.~\ref{fig:akurilkin_data200} by the black circles. 
The error of the experimental values includes both the statistical and systematic errors. The systematic error was derived from the errors of the beam polarization measurements.

The lines in Fig.~\ref{fig:akurilkin_data200} are the results of the relativistic multiple-scattering-model calculations \cite{48}.
Here the parametrized CD-Bonn deuteron \cite{cdbonn} and triton \cite{49} wave functions were used.	
The dashed curves correspond to the calculations including only ONE terms, while the solid curves correspond to the case taking into account both the ONE and single-scattering (SS) contributions.  
The ONE mechanism dominates at forward- and backward-scattering angles \cite{48}, therefore, 
the observed negative and positive signs of the tensor $A_{yy}$, $A_{xx}$ and $A_{xz}$ analyzing powers at forward and backward scattering angles, respectively, reflect different signs of the $D/S$- wave ratios in the triton and deuteron wave functions.  
Strong disagreement of the experimental data with the theoretical calculations taking into account only ONE mechanisms has been observed at the angles between 15$^\circ$ and  160$^\circ$ in the c.m. frame.
The deviation of the vector analyzing power $A_{y}$ from the zero value also indicates that the mechanisms additional to ONE should be considered.
Note that the similar behavior of analyzing powers is observed for the $\overrightarrow{d}d\rightarrow{}^3Hen$ reaction obtained at 270 MeV \cite{44}, where the experimental data demonstrate strong disagreement with the predictions of ONE at the angles larger than $15^\circ$ in the c.m. frame.

The inclusion of the single-scattering diagrams in addition to the ONE mechanism into theoretical calculations significantly improves the description of the experimental data on the tensor analyzing powers.  
The behavior of the vector $A_y$ analyzing power is also reproduced at forward and backward directions in the c.m. frame, which means that the ONE and SS mechanisms give the main contribution to the $\overrightarrow{d}d\rightarrow{}^3Hp$ reaction at these angles. However, large deviations of the theoretical predictions from the experimental data still remain. 
%\textit{
The effort to improve the description of the experimental data using other parametrizations of the deuteron and ${}^3He$ wave functions was undertaken in Ref. \cite{Ladygina_SPIN2012}. 
%The parameterizations based on the Paris NN potentials have been applied for it. 
The obtained results have demonstrated insufficient quantitative difference between the calculations with CD-Bonn and Paris parametrizations \cite{49} and do not describe the existing peaks between $20^\circ$ and $40^\circ$ of the scattering angles.
%}   
A similar problem in the description of the analyzing powers for $dd$- elastic scattering at 231.8 MeV was found for the approximation based on the lowest order terms in the Born series expansion of the Alt-Grassenberger-Sandhas equations for four nucleons interacting with the CD-Bonn potential \cite{dd231.8}. 
%\textit{
Unfortunately, this model was not applied to describe the polarization observables for the $\overrightarrow{d}d\rightarrow{}^3Hen$ reaction in this energy domain.
%}

The discrepancy between the data and the calculations shown in Fig.~\ref{fig:akurilkin_data200} can be explained by the reaction mechanism which differs from ONE and ONE + SS and/or by incomplete description of the short-range ${}^3H$ spin structure. It is possible that the $\Delta$-isobar excitation in the intermediate state may give some contribution to the tensor analyzing powers.

This possibility is discussed in Ref. \cite{50}, where the $\Delta$ isobar is taken into consideration in the simplest phenomenological model. 
%\begin{figure}[b]
%\includegraphics[width=88mm,height=155mm]{akurilkin_data200}% Here is how to import EPS art
%\caption{\label{fig:akurilkin_data200} Results on the the vector $A_{y}$ and tensor $A_{yy}$, $A_{xx}$, $A_{xz}$ analyzing powers in the $\vec{d}d\rightarrow{}^3Hp$ reaction at 200 MeV. The dashed and solid lines are theoretical calculations taking into account the ONE and ONE+SS contributions. }
%\end{figure}

%\section{Conclusion}
High-precision data have been obtained on the $A_{yy}$, $A_{xx}$, $A_{xz}$ and $A_{y}$ analyzing powers in the $\overrightarrow{d}d\rightarrow{}^3Hp$ reactions sensitive to $^3H$ spin structure at the energy of initial deuterons 200 MeV. 
The ONE calculations using ${}^3H$ and deuteron wave functions derived from CD-Bonn potentials have described the data on the tensor $A_{yy}$, $A_{xx}$, and $A_{xz}$ analyzing powers at forward and backward angles. However, they have failed to reproduce strong variations of the tensor analyzing powers as a function of the scattering angle in the c.m. frame. 
The inclusion of the single scattering term into theoretical calculations has significantly improved the description of the experimental data, especially at the backward angles. 
But the behavior of the analyzing powers at forward angles ($15^\circ$--$50^\circ$ in the c.m. frame) has not been reproduced.   
This deviation indicates that further development of theoretical approaches is required. 

%The discrepancy between the data and theoretical calculations indicates either the importance of other mechanisms or the non complete description of the short-range $^3H$ spin structure. 
% Further development of the theoretical approaches to improve the description of the obtained experimental data  is required.

%\begin{acknowledgments}
The authors express their thanks to the staff of RARF for having provided excellent conditions for the R308n experiment. 
Deep recognition and appreciation are expressed to H. Kumasaka, R. Suzuki, and R. Taki for their help during the experiment. 
The Russian part of the collaboration thanks the RIKEN Directorate for kind hospitality during their stay in Japan.
The investigation has been partly supported by a Grant-in-Aid for Scientific Research(Grant $ No.$ 14740151) of the Ministry of Education, Culture, Sports, Science, and Technology of Japan; the Russian Foundation for Basic Research (Grants $ No.$ 10-02-00087a and $ No.$ 13-02-00101a  ).
%\end{acknowledgments}

\bibliography{akurilkin_r308_dd_200}

\providecommand{\noopsort}[1]{}\providecommand{\singleletter}[1]{#1}%
\begin{thebibliography}{38}%
\makeatletter
\providecommand \@ifxundefined [1]{%
 \@ifx{#1\undefined}
}%
\providecommand \@ifnum [1]{%
 \ifnum #1\expandafter \@firstoftwo
 \else \expandafter \@secondoftwo
 \fi
}%
\providecommand \@ifx [1]{%
 \ifx #1\expandafter \@firstoftwo
 \else \expandafter \@secondoftwo
 \fi
}%
\providecommand \natexlab [1]{#1}%
\providecommand \enquote  [1]{``#1''}%
\providecommand \bibnamefont  [1]{#1}%
\providecommand \bibfnamefont [1]{#1}%
\providecommand \citenamefont [1]{#1}%
\providecommand \href@noop [0]{\@secondoftwo}%
\providecommand \href [0]{\begingroup \@sanitize@url \@href}%
\providecommand \@href[1]{\@@startlink{#1}\@@href}%
\providecommand \@@href[1]{\endgroup#1\@@endlink}%
\providecommand \@sanitize@url [0]{\catcode `\\12\catcode `\$12\catcode
  `\&12\catcode `\#12\catcode `\^12\catcode `\_12\catcode `\%12\relax}%
\providecommand \@@startlink[1]{}%
\providecommand \@@endlink[0]{}%
\providecommand \url  [0]{\begingroup\@sanitize@url \@url }%
\providecommand \@url [1]{\endgroup\@href {#1}{\urlprefix }}%
\providecommand \urlprefix  [0]{URL }%
\providecommand \Eprint [0]{\href }%
\providecommand \doibase [0]{http://dx.doi.org/}%
\providecommand \selectlanguage [0]{\@gobble}%
\providecommand \bibinfo  [0]{\@secondoftwo}%
\providecommand \bibfield  [0]{\@secondoftwo}%
\providecommand \translation [1]{[#1]}%
\providecommand \BibitemOpen [0]{}%
\providecommand \bibitemStop [0]{}%
\providecommand \bibitemNoStop [0]{.\EOS\space}%
\providecommand \EOS [0]{\spacefactor3000\relax}%
\providecommand \BibitemShut  [1]{\csname bibitem#1\endcsname}%
\let\auto@bib@innerbib\@empty
%</preamble>
\bibitem [{\citenamefont {Sekiguchi}\ \emph {et~al.}(2005)\citenamefont
  {Sekiguchi} \emph {et~al.}}]{5}%
  \BibitemOpen
  \bibfield  {author} {\bibinfo {author} {\bibfnamefont {K.}~\bibnamefont
  {Sekiguchi}} \emph {et~al.},\ }\href@noop {} {\bibfield  {journal} {\bibinfo
  {journal} {Phys.\ Rev.\ Lett.}\ }\textbf {\bibinfo {volume} {95}},\ \bibinfo
  {pages} {162301} (\bibinfo {year} {2005})}\BibitemShut {NoStop}%
\bibitem [{\citenamefont {Shimizu}\ \emph {et~al.}(1982)\citenamefont {Shimizu}
  \emph {et~al.}}]{11}%
  \BibitemOpen
  \bibfield  {author} {\bibinfo {author} {\bibfnamefont {H.}~\bibnamefont
  {Shimizu}} \emph {et~al.},\ }\href@noop {} {\bibfield  {journal} {\bibinfo
  {journal} {Nucl. Phys. A}\ }\textbf {\bibinfo {volume} {382}},\ \bibinfo
  {pages} {242} (\bibinfo {year} {1982})}\BibitemShut {NoStop}%
\bibitem [{\citenamefont {H.Ruhl}\ \emph {et~al.}(1991)\citenamefont {H.Ruhl}
  \emph {et~al.}}]{12}%
  \BibitemOpen
  \bibfield  {author} {\bibinfo {author} {\bibnamefont {H.Ruhl}} \emph
  {et~al.},\ }\href@noop {} {\bibfield  {journal} {\bibinfo  {journal} {Nucl.
  Phys. A}\ }\textbf {\bibinfo {volume} {524}},\ \bibinfo {pages} {377}
  (\bibinfo {year} {1991})}\BibitemShut {NoStop}%
\bibitem [{\citenamefont {H.Sakai}\ \emph {et~al.}(2000)\citenamefont {H.Sakai}
  \emph {et~al.}}]{13}%
  \BibitemOpen
  \bibfield  {author} {\bibinfo {author} {\bibnamefont {H.Sakai}} \emph
  {et~al.},\ }\href@noop {} {\bibfield  {journal} {\bibinfo  {journal} {Phys.
  Rev. Lett}\ }\textbf {\bibinfo {volume} {84}},\ \bibinfo {pages} {5288}
  (\bibinfo {year} {2000})}\BibitemShut {NoStop}%
\bibitem [{\citenamefont {N.Sakamoto}\ \emph {et~al.}(1996)\citenamefont
  {N.Sakamoto} \emph {et~al.}}]{14}%
  \BibitemOpen
  \bibfield  {author} {\bibinfo {author} {\bibnamefont {N.Sakamoto}} \emph
  {et~al.},\ }\href@noop {} {\bibfield  {journal} {\bibinfo  {journal} {Phys.
  Lett. B}\ }\textbf {\bibinfo {volume} {367}},\ \bibinfo {pages} {60}
  (\bibinfo {year} {1996})}\BibitemShut {NoStop}%
\bibitem [{\citenamefont {K.Sekiguchi}\ \emph {et~al.}(2002)\citenamefont
  {K.Sekiguchi} \emph {et~al.}}]{15}%
  \BibitemOpen
  \bibfield  {author} {\bibinfo {author} {\bibnamefont {K.Sekiguchi}} \emph
  {et~al.},\ }\href@noop {} {\bibfield  {journal} {\bibinfo  {journal} {Phys.
  Rev. C}\ }\textbf {\bibinfo {volume} {65}},\ \bibinfo {pages} {034003}
  (\bibinfo {year} {2002})}\BibitemShut {NoStop}%
\bibitem [{\citenamefont {K.Ermisch}\ \emph {et~al.}(2001)\citenamefont
  {K.Ermisch} \emph {et~al.}}]{16}%
  \BibitemOpen
  \bibfield  {author} {\bibinfo {author} {\bibnamefont {K.Ermisch}} \emph
  {et~al.},\ }\href@noop {} {\bibfield  {journal} {\bibinfo  {journal} {Phys.
  Rev. Lett.}\ }\textbf {\bibinfo {volume} {86}},\ \bibinfo {pages} {5862}
  (\bibinfo {year} {2001})}\BibitemShut {NoStop}%
\bibitem [{\citenamefont {R.Bieber}\ \emph {et~al.}(2000)\citenamefont
  {R.Bieber} \emph {et~al.}}]{17}%
  \BibitemOpen
  \bibfield  {author} {\bibinfo {author} {\bibnamefont {R.Bieber}} \emph
  {et~al.},\ }\href@noop {} {\bibfield  {journal} {\bibinfo  {journal} {Phys.
  Rev. Lett.}\ }\textbf {\bibinfo {volume} {84}},\ \bibinfo {pages} {606}
  (\bibinfo {year} {2000})}\BibitemShut {NoStop}%
\bibitem [{\citenamefont {H.Witala}\ \emph {et~al.}(1993)\citenamefont
  {H.Witala} \emph {et~al.}}]{18}%
  \BibitemOpen
  \bibfield  {author} {\bibinfo {author} {\bibnamefont {H.Witala}} \emph
  {et~al.},\ }\href@noop {} {\bibfield  {journal} {\bibinfo  {journal}
  {Few-Body Syst.}\ }\textbf {\bibinfo {volume} {15}},\ \bibinfo {pages} {67}
  (\bibinfo {year} {1993})}\BibitemShut {NoStop}%
\bibitem [{\citenamefont {J.Arvieux}\ \emph {et~al.}(1983)\citenamefont
  {J.Arvieux} \emph {et~al.}}]{19}%
  \BibitemOpen
  \bibfield  {author} {\bibinfo {author} {\bibnamefont {J.Arvieux}} \emph
  {et~al.},\ }\href@noop {} {\bibfield  {journal} {\bibinfo  {journal} {Phys.
  Rev. Lett.}\ }\textbf {\bibinfo {volume} {50}},\ \bibinfo {pages} {19}
  (\bibinfo {year} {1983})}\BibitemShut {NoStop}%
\bibitem [{\citenamefont {E.J.Stephenson}\ \emph {et~al.}(1999)\citenamefont
  {E.J.Stephenson}, \citenamefont {Witala}, \citenamefont {Glockle},
  \citenamefont {Kamada},\ and\ \citenamefont {A.Nogga}}]{20}%
  \BibitemOpen
  \bibfield  {author} {\bibinfo {author} {\bibnamefont {E.J.Stephenson}},
  \bibinfo {author} {\bibfnamefont {H.}~\bibnamefont {Witala}}, \bibinfo
  {author} {\bibfnamefont {W.}~\bibnamefont {Glockle}}, \bibinfo {author}
  {\bibfnamefont {H.}~\bibnamefont {Kamada}}, \ and\ \bibinfo {author}
  {\bibnamefont {A.Nogga}},\ }\href@noop {} {\bibfield  {journal} {\bibinfo
  {journal} {Phys. Rev. C}\ }\textbf {\bibinfo {volume} {60}},\ \bibinfo
  {pages} {061001(R)} (\bibinfo {year} {1999})}\BibitemShut {NoStop}%
\bibitem [{\citenamefont {R.V.Cadman}\ \emph {et~al.}(2001)\citenamefont
  {R.V.Cadman} \emph {et~al.}}]{21}%
  \BibitemOpen
  \bibfield  {author} {\bibinfo {author} {\bibnamefont {R.V.Cadman}} \emph
  {et~al.},\ }\href@noop {} {\bibfield  {journal} {\bibinfo  {journal} {Phys.
  Rev. Lett.}\ }\textbf {\bibinfo {volume} {86}},\ \bibinfo {pages} {967}
  (\bibinfo {year} {2001})}\BibitemShut {NoStop}%
\bibitem [{\citenamefont {K.Sekiguchi}\ \emph {et~al.}(2004)\citenamefont
  {K.Sekiguchi} \emph {et~al.}}]{22}%
  \BibitemOpen
  \bibfield  {author} {\bibinfo {author} {\bibnamefont {K.Sekiguchi}} \emph
  {et~al.},\ }\href@noop {} {\bibfield  {journal} {\bibinfo  {journal} {Phys.
  Rev. C}\ }\textbf {\bibinfo {volume} {70}},\ \bibinfo {pages} {014001}
  (\bibinfo {year} {2004})}\BibitemShut {NoStop}%
\bibitem [{\citenamefont {K.Hatanaka}\ \emph {et~al.}(2002)\citenamefont
  {K.Hatanaka} \emph {et~al.}}]{23}%
  \BibitemOpen
  \bibfield  {author} {\bibinfo {author} {\bibnamefont {K.Hatanaka}} \emph
  {et~al.},\ }\href@noop {} {\bibfield  {journal} {\bibinfo  {journal} {Phys.
  Rev. C}\ }\textbf {\bibinfo {volume} {66}},\ \bibinfo {pages} {044002}
  (\bibinfo {year} {2002})}\BibitemShut {NoStop}%
\bibitem [{\citenamefont {H.Witala}\ \emph {et~al.}(1998)\citenamefont
  {H.Witala}, \citenamefont {W.Glockle}, \citenamefont {D.Huber}, \citenamefont
  {J.Golak},\ and\ \citenamefont {H.Kamada}}]{24}%
  \BibitemOpen
  \bibfield  {author} {\bibinfo {author} {\bibnamefont {H.Witala}}, \bibinfo
  {author} {\bibnamefont {W.Glockle}}, \bibinfo {author} {\bibnamefont
  {D.Huber}}, \bibinfo {author} {\bibnamefont {J.Golak}}, \ and\ \bibinfo
  {author} {\bibnamefont {H.Kamada}},\ }\href@noop {} {\bibfield  {journal}
  {\bibinfo  {journal} {Phys. Rev. Lett.}\ }\textbf {\bibinfo {volume} {81}},\
  \bibinfo {pages} {1183} (\bibinfo {year} {1998})}\BibitemShut {NoStop}%
\bibitem [{\citenamefont {S.Coon}\ \emph {et~al.}(1979)\citenamefont {S.Coon},
  \citenamefont {M.Scadron}, \citenamefont {P.McNamee}, \citenamefont
  {B.R.Barrett}, \citenamefont {D.Blatt},\ and\ \citenamefont
  {B.McKellar}}]{TM}%
  \BibitemOpen
  \bibfield  {author} {\bibinfo {author} {\bibnamefont {S.Coon}}, \bibinfo
  {author} {\bibnamefont {M.Scadron}}, \bibinfo {author} {\bibnamefont
  {P.McNamee}}, \bibinfo {author} {\bibnamefont {B.R.Barrett}}, \bibinfo
  {author} {\bibnamefont {D.Blatt}}, \ and\ \bibinfo {author} {\bibnamefont
  {B.McKellar}},\ }\href@noop {} {\bibfield  {journal} {\bibinfo  {journal}
  {Nucl. Phys.}\ }\textbf {\bibinfo {volume} {A317}},\ \bibinfo {pages} {242}
  (\bibinfo {year} {1979})}\BibitemShut {NoStop}%
\bibitem [{\citenamefont {B.S.Pudliner}\ \emph {et~al.}(1997)\citenamefont
  {B.S.Pudliner}, \citenamefont {V.R.Pandharipande}, \citenamefont {J.Carlson},
  \citenamefont {S.C.Pieper},\ and\ \citenamefont {Wiringa}}]{UR9}%
  \BibitemOpen
  \bibfield  {author} {\bibinfo {author} {\bibnamefont {B.S.Pudliner}},
  \bibinfo {author} {\bibnamefont {V.R.Pandharipande}}, \bibinfo {author}
  {\bibnamefont {J.Carlson}}, \bibinfo {author} {\bibnamefont {S.C.Pieper}}, \
  and\ \bibinfo {author} {\bibfnamefont {R.~B.}\ \bibnamefont {Wiringa}},\
  }\href@noop {} {\bibfield  {journal} {\bibinfo  {journal} {Phys. Rev. C}\
  }\textbf {\bibinfo {volume} {56}},\ \bibinfo {pages} {1720} (\bibinfo {year}
  {1997})}\BibitemShut {NoStop}%
\bibitem [{\citenamefont {Y.Maeda}\ \emph {et~al.}(2006)\citenamefont {Y.Maeda}
  \emph {et~al.}}]{25}%
  \BibitemOpen
  \bibfield  {author} {\bibinfo {author} {\bibnamefont {Y.Maeda}} \emph
  {et~al.},\ }\href@noop {} {\bibfield  {journal} {\bibinfo  {journal} {AIP
  Conf. Proc.}\ }\textbf {\bibinfo {volume} {915}},\ \bibinfo {pages} {781}
  (\bibinfo {year} {2006})}\BibitemShut {NoStop}%
\bibitem [{\citenamefont {Y.Maeda}\ \emph {et~al.}(2007)\citenamefont {Y.Maeda}
  \emph {et~al.}}]{26}%
  \BibitemOpen
  \bibfield  {author} {\bibinfo {author} {\bibnamefont {Y.Maeda}} \emph
  {et~al.},\ }\href@noop {} {\bibfield  {journal} {\bibinfo  {journal} {Phys.
  Rev. C}\ }\textbf {\bibinfo {volume} {76}},\ \bibinfo {pages} {014004}
  (\bibinfo {year} {2007})}\BibitemShut {NoStop}%
\bibitem [{\citenamefont {W.R.Falk}(1994)}]{27}%
  \BibitemOpen
  \bibfield  {author} {\bibinfo {author} {\bibnamefont {W.R.Falk}},\
  }\href@noop {} {\bibfield  {journal} {\bibinfo  {journal} {Phys. Rev.}\
  }\textbf {\bibinfo {volume} {50}},\ \bibinfo {pages} {1574} (\bibinfo {year}
  {1994})}\BibitemShut {NoStop}%
\bibitem [{\citenamefont {Epstein}\ \emph {et~al.}(1985)\citenamefont {Epstein}
  \emph {et~al.}}]{triumf}%
  \BibitemOpen
  \bibfield  {author} {\bibinfo {author} {\bibfnamefont {M.~B.}\ \bibnamefont
  {Epstein}} \emph {et~al.},\ }\href@noop {} {\bibfield  {journal} {\bibinfo
  {journal} {Phys. Rev. C}\ }\textbf {\bibinfo {volume} {32}},\ \bibinfo
  {pages} {967} (\bibinfo {year} {1985})}\BibitemShut {NoStop}%
\bibitem [{\citenamefont {M.A.Miller}\ \emph {et~al.}(1995)\citenamefont
  {M.A.Miller} \emph {et~al.}}]{37}%
  \BibitemOpen
  \bibfield  {author} {\bibinfo {author} {\bibnamefont {M.A.Miller}} \emph
  {et~al.},\ }\href@noop {} {\bibfield  {journal} {\bibinfo  {journal} {Phys.
  Rev. Lett}\ }\textbf {\bibinfo {volume} {74}},\ \bibinfo {pages} {502}
  (\bibinfo {year} {1995})}\BibitemShut {NoStop}%
\bibitem [{\citenamefont {B.Blankleider}\ and\ \citenamefont
  {R.M.Woloshyn}(1984)}]{38}%
  \BibitemOpen
  \bibfield  {author} {\bibinfo {author} {\bibnamefont {B.Blankleider}}\ and\
  \bibinfo {author} {\bibnamefont {R.M.Woloshyn}},\ }\href@noop {} {\bibfield
  {journal} {\bibinfo  {journal} {Phys. Rev. C}\ }\textbf {\bibinfo {volume}
  {29}},\ \bibinfo {pages} {538} (\bibinfo {year} {1984})}\BibitemShut
  {NoStop}%
\bibitem [{\citenamefont {Shimizu}\ \emph {et~al.}(2007)\citenamefont {Shimizu}
  \emph {et~al.}}]{41}%
  \BibitemOpen
  \bibfield  {author} {\bibinfo {author} {\bibfnamefont {Y.}~\bibnamefont
  {Shimizu}} \emph {et~al.},\ }\href@noop {} {\bibfield  {journal} {\bibinfo
  {journal} {Phys. Rev. C}\ }\textbf {\bibinfo {volume} {76}},\ \bibinfo
  {pages} {044003} (\bibinfo {year} {2007})}\BibitemShut {NoStop}%
\bibitem [{\citenamefont {V.P.Ladygin}\ and\ \citenamefont
  {N.B.Ladygina}(2002)}]{42}%
  \BibitemOpen
  \bibfield  {author} {\bibinfo {author} {\bibnamefont {V.P.Ladygin}}\ and\
  \bibinfo {author} {\bibnamefont {N.B.Ladygina}},\ }\href@noop {} {\bibfield
  {journal} {\bibinfo  {journal} {Phys. Atom. Nucl.}\ }\textbf {\bibinfo
  {volume} {65}},\ \bibinfo {pages} {1609} (\bibinfo {year}
  {2002})}\BibitemShut {NoStop}%
\bibitem [{\citenamefont {V.P.Ladygin}\ \emph {et~al.}(2004)\citenamefont
  {V.P.Ladygin} \emph {et~al.}}]{43}%
  \BibitemOpen
  \bibfield  {author} {\bibinfo {author} {\bibnamefont {V.P.Ladygin}} \emph
  {et~al.},\ }\href@noop {} {\bibfield  {journal} {\bibinfo  {journal} {Phys.
  Lett.B}\ }\textbf {\bibinfo {volume} {589}},\ \bibinfo {pages} {47} (\bibinfo
  {year} {2004})}\BibitemShut {NoStop}%
\bibitem [{\citenamefont {Janek}\ \emph {et~al.}(2007)\citenamefont {Janek}
  \emph {et~al.}}]{44}%
  \BibitemOpen
  \bibfield  {author} {\bibinfo {author} {\bibfnamefont {M.}~\bibnamefont
  {Janek}} \emph {et~al.},\ }\href@noop {} {\bibfield  {journal} {\bibinfo
  {journal} {Eur.Phys.J.}\ }\textbf {\bibinfo {volume} {A33}},\ \bibinfo
  {pages} {39} (\bibinfo {year} {2007})}\BibitemShut {NoStop}%
\bibitem [{\citenamefont {P.Grassberger}\ and\ \citenamefont
  {W.Sandhas}(1967)}]{ags}%
  \BibitemOpen
  \bibfield  {author} {\bibinfo {author} {\bibnamefont {P.Grassberger}}\ and\
  \bibinfo {author} {\bibnamefont {W.Sandhas}},\ }\href@noop {} {\bibfield
  {journal} {\bibinfo  {journal} {Nucl.Phys.}\ }\textbf {\bibinfo {volume}
  {B2}},\ \bibinfo {pages} {181} (\bibinfo {year} {1967})}\BibitemShut
  {NoStop}%
\bibitem [{\citenamefont {N.B.Ladygina}(2012)}]{48}%
  \BibitemOpen
  \bibfield  {author} {\bibinfo {author} {\bibnamefont {N.B.Ladygina}},\
  }\href@noop {} {\bibfield  {journal} {\bibinfo  {journal} {Few Body Syst.}\
  }\textbf {\bibinfo {volume} {53}},\ \bibinfo {pages} {253} (\bibinfo {year}
  {2012})}\BibitemShut {NoStop}%
\bibitem [{\citenamefont {Ladygina}(2012{\natexlab{a}})}]{Ladygina_ISHEPP2012}%
  \BibitemOpen
  \bibfield  {author} {\bibinfo {author} {\bibfnamefont {N.}~\bibnamefont
  {Ladygina}},\ }\href@noop {} {\enquote {\bibinfo {title} {{Proceedings of the
  XXI International Baldin Seminar on High Energy Physics Problems, Dubna,
  Russia}},}\ } (\bibinfo {year} {2012}{\natexlab{a}}),\ \bibinfo {note} {(to
  be published in PoS)}\BibitemShut {NoStop}%
\bibitem [{\citenamefont {Okamura}\ \emph {et~al.}(1993)\citenamefont {Okamura}
  \emph {et~al.}}]{45}%
  \BibitemOpen
  \bibfield  {author} {\bibinfo {author} {\bibfnamefont {H.}~\bibnamefont
  {Okamura}} \emph {et~al.},\ }\href@noop {} {\bibfield  {journal} {\bibinfo
  {journal} {AIP Conf. Proc.}\ }\textbf {\bibinfo {volume} {293}},\ \bibinfo
  {pages} {84} (\bibinfo {year} {1993})}\BibitemShut {NoStop}%
\bibitem [{\citenamefont {Ichihara}\ \emph {et~al.}(1994)\citenamefont
  {Ichihara} \emph {et~al.}}]{46}%
  \BibitemOpen
  \bibfield  {author} {\bibinfo {author} {\bibfnamefont {T.}~\bibnamefont
  {Ichihara}} \emph {et~al.},\ }\href@noop {} {\bibfield  {journal} {\bibinfo
  {journal} {Nucl. Phys. A}\ }\textbf {\bibinfo {volume} {569}},\ \bibinfo
  {pages} {287} (\bibinfo {year} {1994})}\BibitemShut {NoStop}%
\bibitem [{\citenamefont {Maeda}\ \emph {et~al.}(2002)\citenamefont {Maeda},
  \citenamefont {Sakai}, \citenamefont {Hatanaka},\ and\ \citenamefont
  {Tamii}}]{47}%
  \BibitemOpen
  \bibfield  {author} {\bibinfo {author} {\bibfnamefont {Y.}~\bibnamefont
  {Maeda}}, \bibinfo {author} {\bibfnamefont {H.}~\bibnamefont {Sakai}},
  \bibinfo {author} {\bibfnamefont {K.}~\bibnamefont {Hatanaka}}, \ and\
  \bibinfo {author} {\bibfnamefont {A.}~\bibnamefont {Tamii}},\ }\href@noop {}
  {\bibfield  {journal} {\bibinfo  {journal} {Nucl.Instrum. Methods in
  Phys.Res. A}\ }\textbf {\bibinfo {volume} {490}},\ \bibinfo {pages} {518}
  (\bibinfo {year} {2002})}\BibitemShut {NoStop}%
\bibitem [{\citenamefont {Machleidt}(2001)}]{cdbonn}%
  \BibitemOpen
  \bibfield  {author} {\bibinfo {author} {\bibfnamefont {R.}~\bibnamefont
  {Machleidt}},\ }\href@noop {} {\bibfield  {journal} {\bibinfo  {journal}
  {Phys. Rev. C}\ }\textbf {\bibinfo {volume} {63}},\ \bibinfo {pages} {024001}
  (\bibinfo {year} {2001})}\BibitemShut {NoStop}%
\bibitem [{\citenamefont {Baru}\ \emph {et~al.}(2003)\citenamefont {Baru} \emph
  {et~al.}}]{49}%
  \BibitemOpen
  \bibfield  {author} {\bibinfo {author} {\bibfnamefont {V.}~\bibnamefont
  {Baru}} \emph {et~al.},\ }\href@noop {} {\bibfield  {journal} {\bibinfo
  {journal} {Eur.Phys.J.}\ }\textbf {\bibinfo {volume} {A16}},\ \bibinfo
  {pages} {437} (\bibinfo {year} {2003})}\BibitemShut {NoStop}%
\bibitem [{\citenamefont {Ladygina}(2012{\natexlab{b}})}]{Ladygina_SPIN2012}%
  \BibitemOpen
  \bibfield  {author} {\bibinfo {author} {\bibfnamefont {N.}~\bibnamefont
  {Ladygina}},\ }\href@noop {} {\enquote {\bibinfo {title} {{Proceedings of XX
  the International Symposium on Spin Physics, SPIN 2012, Dubna, Russia}},}\ }
  (\bibinfo {year} {2012}{\natexlab{b}}),\ \bibinfo {note} {(to be
  published)}\BibitemShut {NoStop}%
\bibitem [{\citenamefont {A.M.Micherdzi\'nska}\ \emph
  {et~al.}(2007)\citenamefont {A.M.Micherdzi\'nska} \emph {et~al.}}]{dd231.8}%
  \BibitemOpen
  \bibfield  {author} {\bibinfo {author} {\bibnamefont {A.M.Micherdzi\'nska}}
  \emph {et~al.},\ }\href@noop {} {\bibfield  {journal} {\bibinfo  {journal}
  {Phys. Rev. C}\ }\textbf {\bibinfo {volume} {75}},\ \bibinfo {pages} {054001}
  (\bibinfo {year} {2007})}\BibitemShut {NoStop}%
\bibitem [{\citenamefont {G.Bizard}\ \emph {et~al.}(1980)\citenamefont
  {G.Bizard} \emph {et~al.}}]{50}%
  \BibitemOpen
  \bibfield  {author} {\bibinfo {author} {\bibnamefont {G.Bizard}} \emph
  {et~al.},\ }\href@noop {} {\bibfield  {journal} {\bibinfo  {journal}
  {Phys.Rev.C}\ }\textbf {\bibinfo {volume} {22}},\ \bibinfo {pages} {1632}
  (\bibinfo {year} {1980})}\BibitemShut {NoStop}%
\end{thebibliography}%

% Produces the bibliography via BibTeX.

\end{document}